\documentstyle[epsf]{mn}

\begin{document}
\title[XTE J0111.2--7317]
    {The SMC X-ray transient XTE J0111.2--7317 : a Be/X-ray binary in
a SNR?}

\author[M.J.Coe et al.]
{M.J. Coe$^{1}$, N.J. Haigh$^{1}$, P. Reig$^{2,3}$ \\
$^{1}$Physics and Astronomy Dept., The University, Southampton, SO17
3BJ, UK. \\
$^{2}$Foundation for Research and Technology-Hellas, 711 10
Heraklion, Crete, Greece.\\
$^{3}$Physics Department, University of Crete, 71003 Heraklion, Crete, Greece.\\
}

\date{Accepted \\
Received : Version 19 Nov 1999 \\
In original form ..}

\maketitle

\begin{abstract} 

We report observations which confirm the identity of the optical/IR
counterpart to the Rossi X-ray Timing Explorer transient source XTE
J0111.2--7317.  The counterpart is suggested to be a B0-B2 star
(luminosity class III--V) showing an IR excess and strong Balmer
emission lines. The distance derived from reddening and systemic
velocity measurements puts the source in the SMC. Unusually, the
source exhibits an extended asymetric H$\alpha$ structure.

\end{abstract}


\section{Introduction}

The Be/X-ray and supergiant binary systems represent the
class of massive X-ray binaries.  A survey of the literature
reveals that of the 96 proposed massive X-ray binary pulsar systems,
67\% of the identified systems fall within 
the Be/X-ray group of binaries.  The orbit
of the Be star and the compact object, presumably a
neutron star, is generally wide and eccentric.  The optical star
exhibits H$\alpha$ line emission and continuum free-free emission
(revealed as excess flux in the IR) from a disk of circumstellar
gas. Most of the Be/X-ray sources are also very transient in the
emission of X-rays.
 
Progress towards a better understanding of the physics of these systems
depends on a multi-wavelength programme of observations.  From
observations of the Be star in the optical and IR, the physical
conditions under which the neutron star is accreting matter can be
determined.  In combination with hard X-ray timing and flux
observations, this yields a near complete picture of the accretion
process.  It is thus vital to identify the optical counterparts to
these X-ray systems in order to further our understanding. 

The X-ray transient XTE J0111.2--7371 was discovered by the RXTE X-ray
observatory in November 1998 (Chakrabarty et al. 1998a,b) as a 31s
X-ray pulsar. This detection was confirmed by observations from the
BATSE telescope on the CGRO spacecraft which detected the source in
the 20--50 keV band. The quick-look results provided by the ASM/RXTE
team indicate that the X-ray source was active for the period November
1998 -- January 1999. A Be star counterpart has been
proposed by Israel et al. (1999) based upon optical spectroscopy.


Reported here are optical and infra-red measurements taken while the
source was X-ray active which confirm the proposed identity of the
counterpart to XTE J0111.2--7317. The counterpart is shown to be most
consistent with a main sequence B0--B2 star. In addition, strong
evidence is presented from the H$\alpha$ imaging for a surrounding
nebula, possibly a SNR.

\section{Optical spectroscopy}

Optical spectra were taken on 9-10 January 1999 from the SAAO 1.9m
telescope. The instrument used was the grating spectrograph with the
SITe CCD detector. See Table 1 for details of the observation configurations.

\begin{table}
 \centering
 \caption{Journal of spectroscopic observations}
 \begin{tabular}{cccc}

Date & UTC & Wavelength & Dispersion \\
&& Range (\AA)& (\AA/pixel)\\
&&&\\
1999 Jan 9&21.15&6250-6900&0.75 \\
1996 Jan 9&21.53&6250-6900&0.75 \\
1996 Jan 10&20.44&3700-5400&1.5 \\

\end{tabular}
\end{table}

The only significant lines detected were from strong H$\alpha$ and H$\beta$
emission. The equivalent widths obtained were $-27\pm$0.3\AA ~for
H$\alpha$ and $-3.8\pm$0.2\AA ~for H$\beta$. Furthermore the average
redshift of the lines from their rest position was measured and this 
corresponded to a recessional velocity of 165$\pm$15 km/s.

The other bright star in the error circle was also checked but did not
show any evidence of H$\alpha$ in emission.

\section{Optical and IR photometry}

Photometry of the source was obtained from South Africa using the 1.9m
and the 1.0m telescopes in January 1999. The 1.9m IR data were
collected using the Mk III photometer in the J and H bands. The 1.0m
data were obtained using the Tek8 CCD, giving a field of approximately
3 arcminutes, and a pixel scale of 0.3'' per pixel. Observations were
made through standard Johnson UBVRI and Str{\"o}mgren-Crawford
uvby$\beta$ filters.  The exact dates and filters used are specified
in Table 2.  The 1.0m data were reduced using IRAF and Starlink
software, and the instrumental magnitudes were corrected to the
standard system using E region standards.

Figure 1 shows a V band image from our observations which shows the
X-ray uncertainty (30'' radius error circle) and the candidate
proposed by Israel et al., 1999.

\begin{figure}
\begin{center}
{\epsfxsize 0.99\hsize
 \leavevmode
 \epsffile{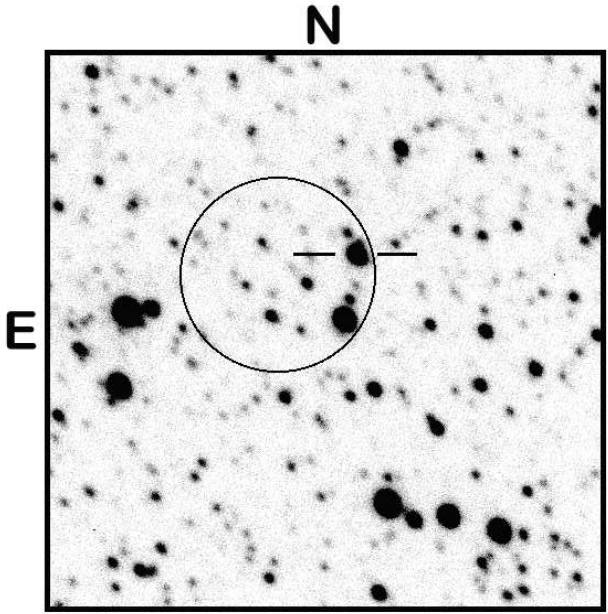}
}\end{center}
\caption{A V band image of the field containing the source XTE
J0111.2--7317 (marked) and the X-ray error circle (30'' radius) 
from the RXTE satellite. The full
field of view is approximately 3 x 3 arcminutes.}
\label{}
\end{figure}

\begin{table}
 \centering
 \caption{Optical and IR photometry of the counterpart to XTE
J0111.2--7317.}
 \begin{tabular}{lcc}

Band & 8 Jan 1999&21 Jan 1999\\
&&\\
B&& 15.24$\pm$0.01\\
V&& 15.32$\pm$0.01\\
R&& 15.37$\pm$0.02\\
I&& 15.31$\pm$0.01\\
u&& 15.90$\pm$0.02\\
v&& 15.56$\pm$0.02\\
b&& 15.53$\pm$0.01\\
y&& 15.44$\pm$0.02\\
J&15.5$\pm$0.3&\\
H&15.1$\pm$0.2&\\

 \end{tabular} 
\end{table} 

The observed $\beta$ index was determined by taking the
ratio of the fluxes in the H$\beta$-wide and the H$\beta$-narrow
filters. This was found to be 2.38 before any correction for the
circumstellar emission was applied (see Section 5.1 for a further
discussion of this point). 

\section{H$\alpha$ imaging}

Of particular interest with this source is the H$\alpha$ image. A
1000s exposure taken on 21 January 1999 with the 1.0m telescope 
revealed a clear extended
structure around the Be star. Consequently a deeper 2000s image was
recorded on 24 January 1999.
This latter image is shown in Figure 2. Also shown in the figure is an
(H$\alpha$--R) image. This latter image was obtained by normalising
the two separate images, registering the fields to sub-pixel accuracy,
and then subtracting one from the other. Despite the accurate
registration, variations in the PSF have caused some residual
structures. Nonetheless, the result is an
image which clearly shows the sources of H$\alpha$ within the
field. In addition to the extended structure around XTE J0111.2--7317
and its associated Be star, one can see two other strong stellar
H$\alpha$ emitters at the eastern and southern edges of the
field. These are probably also both Be stars.

\begin{figure}
\begin{center}
{\epsfxsize 0.99\hsize
 \leavevmode
 \epsffile{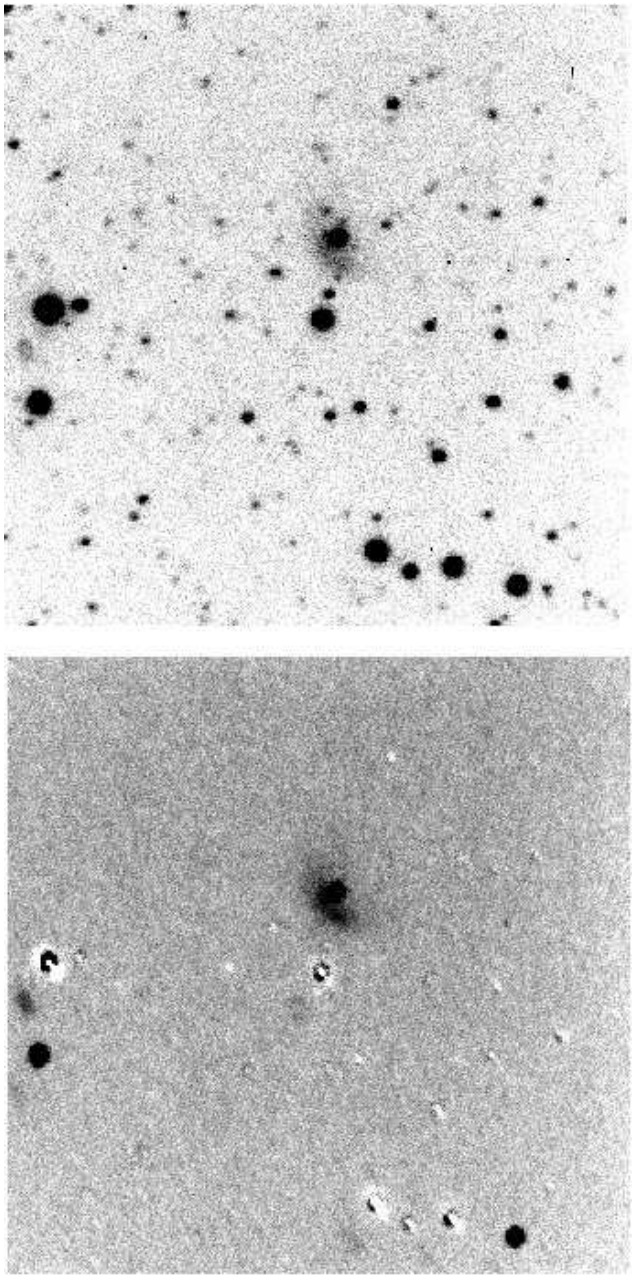}
}\end{center}
\caption{Upper panel : a deep H$\alpha$ image of 
the region around the proposed
counterpart to XTE J0111.2--7317. Lower panel : the same field in
which the R band image has been subtracted from the H$\alpha$ image.
Both images are the same field size and orientation as that of
Figure 1.}
\label{}
\end{figure}

\section{Discussion}

\subsection{Spectral Class}

The equivalent width obtained for the target was $-27\pm$0.3\AA ~for
H$\alpha$. Typical equivalent widths found in Be stars are in the
range 0 to -40\AA, while those found in supergiants stars lie below
-4\AA. Some hypergiants (luminosity class Ia+) can reach
-7\AA (eg Wray977, Kaper et al, 1995), but none have been reported
greater than -10\AA. Thus the high H$\alpha$ value is a strong
indicator that this source is a Be star. 

The determination of the spectral type and luminosity class in a Be star
is not as straightforward as in a non-emission line B-type star due to the
presence of the surrounding envelope, which distorts the characteristic
photospheric spectrum.  In the $(b-y)_0$ --$M_V$ plane Be stars appear
redder than the non emission B stars, due to the additional reddening
caused by the hydrogen free-bound and free-free recombination in the
circumstellar envelope. In the $c_0$--$M_V$ plane the earlier Be stars
present lower values than absorption-line B stars, which is caused by
emission in the Balmer discontinuity, while the later Be stars deviate
towards higher values, indicating absorption in the Balmer discontinuity
of circumstellar origin (Fabregat et al. 1996). 

Thus, in a Be star one has to correct for both circumstellar and
interstellar reddening before any calibration can be used. There is no
easy way to decouple these two reddening contributions. The 
iterative procedure of Fabregat \& Torrej\'on (1998) has been used here to
correct for both circumstellar and interstellar reddening. 


The mean values of the interstellar and circumstellar reddening are
$E^{is}(b-y)$=0.104$\pm$0.033 and $E^{cs}(b-y)$=0.091$\pm$0.025,
respectively. The errors reflect the accuracy of the Fabregat \&
Torrej\'on method. From the relation $E(B-V)=1.35 E(b-y)$ (Crawford \&
Mandwewala 1976) we find that the {\em total} (i.e., interstellar plus
circumstellar) extinction is $E(B-V)$=0.26$\pm$0.05. 

In support of the method we point out that the derived mean interstellar
excess colour $E^{is}(B-V)$=0.14$\pm$0.04 is compatible with the average
extinction reported for the SMC. Schwering \& Israel (1991) have measured
the extinction E(B-V) as lying in the range 0.07--0.09, though there is
no doubt that in some localised regions in the SMC the value can be as
high as 0.25.

The $c_0$ index is the primary temperature parameter for O and B type
stars (Crawford 1978). Using the calibration
$logT_{eff}$=$0.186c_0^2-0.580c_0+4.402$ (Reig et al. 1997) we obtain
an effective temperature of 21800$\pm$1200 K. Using the calibration of
Balona (1994) a virtually identical value of 22000 K is
obtained. Such temperature is typical of a B1-B2 star (Zombek
1992). The same spectral class is obtained from the $(b-y)_0$ index,
which is also a temperature indicator. The derived value, --104,
agrees with a B1-B2 main sequence star (Moon 1985).

Another way of determing the spectral type is by means of the Q method.
The Q parameter is defined as Q=(U-B)-0.72(B-V) and it is independent of
reddening. We obtain Q=--0.892, which according to Halbedel (1993)
corresponds to a B1e star, in agreement with the above analysis.

The $\beta$ index (Crawford \& Mander 1966) provides a measure of
luminosity class for O and B type stars. However, it is also strongly
affected by circumstellar emission, with the extra complication that
there is not any other independent index which can be related to the
stellar luminosity. The value of $\beta_0$=2.647 or equivalently
$M_v=-2.2\pm0.7$ (Balona \& Shobbrook 1984) is consistent with a B2V
star (Crawford 1978; Moon 1985). However, the distance implied
from$M_v$ and V is about 3 times lower than the distance to the SMC. A
better approach is to make use of the knowledge of the distance
modulus to the SMC. The apparent V magnitude is
$m_v$=15.32$\pm$0.01. Asuming a distance modulus of
$(m_v-M_v)_0$=18.8$\pm$0.1 (Westerlund 1997, Table 2.4) and the
derived $E^{is}(B-V)=$0.14$\pm$0.04, the absolute magnitude $M_v$
comes out to be $M_v=-3.8\pm0.2$, closer to a B0V than to a B2III
star.

If, for the sake of discussion, we assume that the correct spectral
class is B1V, then the intrinsic B--V for such an object is -0.26. The
photometric measurements presented here have B--V=-0.08$\pm$0.02,
suggesting an E(B--V)=0.18$\pm$0.02, just about consistent with our
above value from the Stromgren photometry of E(B--V)=0.26$\pm$0.05. If
the optical/IR photometry are then dereddened by our estimate of the
interstellar extinction ($E^{is}(B-V)$=0.14) it is possible to compare
the results with the model atmosphere expected for stars in the range
B0-B2 ($T_{eff}$=30000K -- 19000K and log g=4.0). These comparisons
are shown in Figure 3 where the spectra have all been normalised to
the B band flux. The agreement is good over the B -- R range, though
at longer wavelengths the usual infrared excess arising from the
circumstellar disk is clearly present.

\begin{figure}
\begin{center}
{\epsfxsize 0.99\hsize
 \leavevmode
 \epsffile{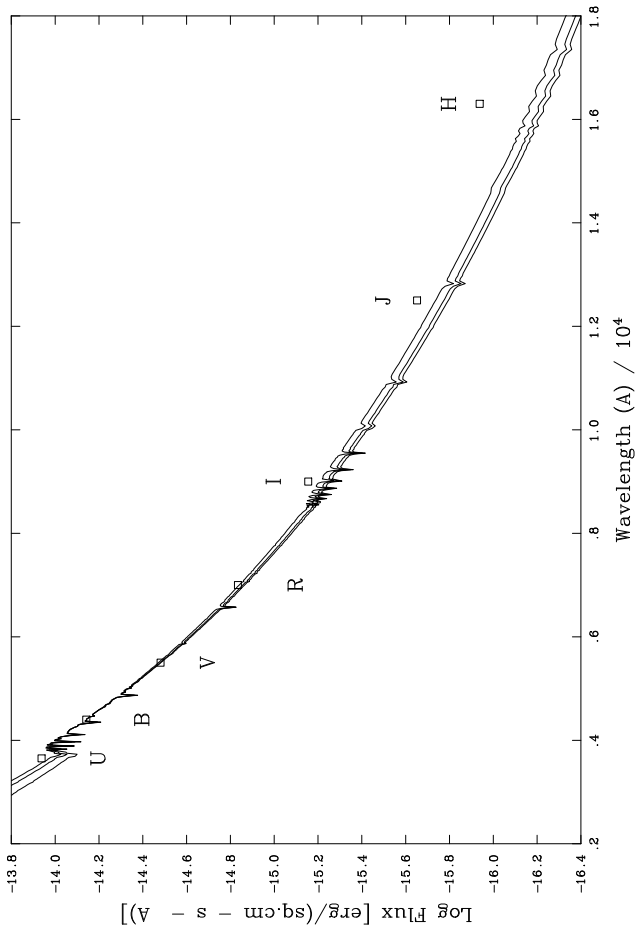}
}\end{center}
\caption{Comparison of the observed photometric fluxes (dereddened by
E(B--V)=0.14) with the model atmosphere expected for B0 -- B2 main
sequence stars.
}
\label{}
\end{figure}

The data presented in this work 
include the first reported IR measurements of this system
and confirm it to be an IR source similar to other Be/X-ray systems.
Our data indicate an apparent colour index of (J--H) $\approx$
0.40$\pm$0.36, very similar to that seen from other SMC X-ray
transients such as RX J0117.6-7330 (Coe et al. 1998).


\subsection{Extended emission}

Two previous Be/X-ray binary systems have been shown by Hughes and
Smith (1994) to exhibit extended H$\alpha$ emission close to, or
directly linked to the Be star. In both those objects the authors
expressed the view that it was unlikely that we were observing the SNR
left over from the production of the neutron star in the Be/X-ray
binary system. This argument was based upon the high transverse speeds
that would be needed to explain the separation of the neutron star
from the deduced centre of the SNR. Such high radial velocities
(approximately 100 km/s) are not seen in the optical lines, though the
inclination of the motion to the line of sight would obviously affect
the amount that was observable.  They therefore thought it more likely
that the SNR was produced from another object in the same OB
association that gave rise to the Be/X-ray system. However, they did
express some disquiet over finding two such systems.

In the case presented here of XTE J0111.2--7317, the association of
the extended structure with the Be star seems much stronger. It is
morphologically much closer than the systems of Hughes and Smith
(1994) and so such high SN kick velocities are not required. One
observational test that will be possible when we have better orbital
data, is to search for the high eccentricity that would inevitably be
associated with a large SN kick. 

An alternative explanation for the extended emission is the presence
of a wind bow shock such as that detected by Kaper et al (1997) from
Vela X-1. In that case the extended H$\alpha$ emssion was explained as
arising from supersonic motion of the system through the interstellar
medium. However, such bow shocks are more likely to occur in
supergiant systems than in a Be star system. More detailed radio and
H$\alpha$ imaging may help resolve the structure of the emission.

In addition to the above 3 sources, Yokogawa et al (1999) have
identified a 4th system in the SMC (AX J0105-722) which appears to be
associated with the the radio SNR DEM S128. So we now have 4 systems
apparently associated with SNR out of the 16 probable Be/X-ray binary
pulsars in the SMC.

\subsection{Distance}

The conclusion that this object is in the SMC 
is supported by the average velocity shift of the
optical lines of 166$\pm$15 km/s. This compares favourably with the
systemic value of 166$\pm$3 km/s obtained by Feast (1961) for the SMC.

\section{Conclusions}

In summary, the optical photometric measurements presented here
indicate that the most likely candidate to the X-ray transient XTE
J0111.2--7317 is a B1$\pm$1 main sequence star. It is noteworthy that
this classification falls within the narrow range of spectral types
in Be/X-ray binaries, namely O9--B2. Optical spectroscopic
observations in the wavelength range 4000--4800 \AA \ are encouraged
in order to refine this classification. Further studies of the
extended structure will also be important in identifying its nature.

\subsection*{Acknowledgments}

We are grateful to the very helpful staff at the SAAO for their
support during these observations.  All of the data reduction was
carried out on the Southampton Starlink node which is funded by the
PPARC.  NJH is in receipt of a PPARC studentship and PR acknowledges
support from the European Union through the Training and Mobility
Research Network Grant ERBFMRX/CT98/0195.

In addition, we gratefully acknowledge helpful contributions from the
referee, Dr. L. Kaper.

\bsp

\end{document}